\documentclass[aps,prb,twocolumn,superscriptaddress,showpacs]{revtex4}
\usepackage{ae}
\usepackage[T1]{fontenc}
\usepackage[ansinew]{inputenc}
\usepackage{amsmath}
\usepackage{amssymb}
\usepackage{graphicx}
\usepackage{color}
\usepackage[colorlinks]{hyperref}
\usepackage{epstopdf}

\begin{document}

\date{\today}

\title{Dynamics of quasi-particles in graphene with impurities and sharp edges
from the \textbf{kp}-method standpoint.}

\author{A.M. Kadigrobov}
\affiliation{Theoretische Physik III, Ruhr-Universitaet Bochum, D-44801 Bochum, Germany}


\begin{abstract}
Dynamics of quasi-particles in graphene with an impurity and a sharp edge is considered
with the \textbf{kp}-method that allows an unified approach  without
usage of any models.
Dirac and Weyl equations are derived by the above-mentioned method. The wave function
 and its  envelope function  together with the scattering
 amplitude are found in the Born approximation. The wave functions are shown to be a superposition of
 virtual Bloch functions which exponential decay outward from the impurity and the edge. At  distances much
 greater that the atomic spacing the wave functions are explicitly presented.
Green's functions for Shr\"{o}dinger and Dirac equations are derived as well.  Boundary conditions for the
 Dirac equation for graphene with a sharp edge are also derived.
\end{abstract}

\maketitle
\section{Introduction.}

Dynamic and  kinetic properties of graphene  have been  attracting much attention
during the last decades \cite{Beenakker}. Fascinating dynamic and kinetic  phenomena  which arise in  graphene
can be described by the two dimensional  differential
Dirac equation \cite{Wallace,Vincenzo} supplemented by boundary conditions.

Details of the boundary conditions and  scattering amplitudes
depend on microscopic characteristics of the concrete
structures of  sample boundaries\cite{Son} and the scatterers.
Theoretical derivations of the boundary conditions  for Dirac equations and the scattering amplitudes are usually based on
various models such as tight bound model
(see, e.g., review papers \cite{Neto,Sarma} and references there), the effective mass model \cite{Falko}, tight-binding model with
a staggered potential at a zigzag boundary \cite{AkhmerovBeenakker}.

The object of this paper is to demonstrate that the   $kp$-method \cite{LL} allows  investigations of graphene (and Weyl semi-metals)
fundamental properties (including the above-mentioned) on the base a unified  approach. This approach is justified by the fact that
the cone points in  graphene  are on the Fermi level or close to it while
 the $kp$-approximation requires  nothing but the series  expansion in the quasi-particle momenta in their vicinity.

In this paper, on the basis of the $kp$-method  and without usage of any models 1) the Dirac and Weyl equations for quasi-particles
are derived;
2) Green's function, the  wave function  together with the scattering amplitude for graphene with an impurity  are obtained in terms
of Bloch functions; 3) Dirac equation, the envelope function together with the scattering amplitude,
 Dirac equation for Green function, Green's function   are found; 4)
 boundary conditions for the Dirac equation for quasi-particles in graphene with an abrupt boundary
 are also presented (details of their derivation in the $kp$-approximation were earlier published in Ref.\cite{scattprobl}).

The outline of this paper is as follows.   In Sec.\ref{SecDiracDerivatuion}
 Dirac  and Weyl equations are obtained on the base of the \textbf{kp}-approximation.
In Sec.\ref{SecImpScatt} elastic scattering of quasi-particles by an impurity   in graphene  is considered:
the wave function, the envelope function
and  the scattering amplitude are found in the Born approximation;
 Green's functions for Schr\"{o}dinger   and  Dirac equations are also obtained.
In Sec.\ref{BoundaryConditions} dynamics of quasi-particles and boundary conditions  at the sharp edge of graphene  is  shortly
described. In Sec.\ref{conclusion} concluding remarks are presented.

\section{Derivation of  Dirac equation by $kp$-method.  \label{SecDiracDerivatuion}}

Here we shortly present derivation of  the Dirac equation by the $kp$-method assuming that two quasi-particle energy bands are degenerated at a point $\mathbf{p}_0=0$ ("Dirac" point) in the quasi-momentum space (see Ref. \cite{scattprobl} for details).


The Schr\"odinger equation for  noninteracting  quasi-particles   is written as
\begin{eqnarray}
\left[-\frac{\hbar^2}{2 m}\frac{\partial^2}{\partial\mathbf{ r}^2} +U(\mathbf{r})\right] \varphi_{s,\mathbf{p}}(\mathbf{r})=\varepsilon_s(\mathbf{p})
\varphi_{s,\mathbf{p}}(\mathbf{r})
 \label{Schroedinger0}
\end{eqnarray}
where $U(\mathbf{r})=U(\mathbf{r}+\mathbf{a})$ is the lattice periodic  potential ($\mathbf{a}$ is
the lattice vector)
and
\begin{eqnarray}
 \varphi_{s,\mathbf{p}}(\mathbf{r})=e^{i\frac{\mathbf{pr}}{\hbar}}u_{s,\mathbf{p}}(\mathbf{r})
 \label{Bloch}
\end{eqnarray}
is the Bloch function and $u_{s,\mathbf{p}}(\mathbf{r})$ is its periodic factor, $\mathbf{p}$ is
the electron quasi-momentum while $\varepsilon_s(\mathbf{p})$ is the dispersion law and $s$ is the band number.

For further application of the $kp$-method it is convenient to re-write the Schr\"odinger equation, Eq(\ref{Schroedinger0}),
as follows:
\begin{eqnarray}
\Big \{ \frac{1}{2m} \left(-i\hbar \frac{\partial }{\partial\mathbf{ r}}+ \mathbf{p} \right)^2 +U(\mathbf{r}) \Big\}
u_{s,\mathbf{p}}(\mathbf{r})=\varepsilon_s(\mathbf{p})u_{s,\mathbf{p}}(\mathbf{r})
 \label{SchroederBF}
\end{eqnarray}

According to the $kp$-method one finds the quasi-particle dispersion law
in the vicinity of the degeneration point
presenting the proper wave functions   as a superposition  of   Lattinger-Kohn functions\cite{LL}
\begin{eqnarray}
\chi_{\alpha,\mathbf{p}}=\exp{\left(i\frac{\mathbf{pr}}{\hbar}\right)}
u_{s,0}(\mathbf{r})
 \label{chi}
\end{eqnarray}
where    the periodic Bloch factors   are taken at the degeneration point $\mathbf{p}=0$.

In order to solve Eq.(\ref{SchroederBF}) by the perturbation theory with degeneration
one takes the sought-for function as a superposition of the degenerated ones that is
\begin{eqnarray}
u_{s,\mathbf{p}}(\mathbf{r})= g_1 (\mathbf{p})u_{1,0}(\mathbf{r}) + g_2(\mathbf{p})
u_{2,0}(\mathbf{r})
 \label{PerturbExpasion}
\end{eqnarray}

Inserting the wave function, Eq.(\ref{PerturbExpasion}), in
Eq.(\ref{SchroederBF}) and using  the
inequality $|\mathbf{p}| \ll  \hbar/a$  one obtains  the former equation in
the following form:
\begin{eqnarray}
\left[-\frac{\hbar^2}{2m} \frac{\partial^2}{\partial \mathbf{r}^2}+U(\mathbf{r}) +\mathbf{p} \hat{\mathbf{v}}  - \varepsilon \right]
\sum_{\alpha =1}^2 g_{\alpha} u_{\alpha,0}(\mathbf{r}) =0;
 \label{PertubSchroedinger}
\end{eqnarray}
Here  $\hat{\mathbf{v}}=(-i\hbar/m)\partial/\partial\mathbf{ r} $  is  the velocity operator,
 $\alpha \equiv s=1,2$ are the band numbers of the two degenerated bands.

Taking  matrix elements of  Eq.(\ref{PertubSchroedinger}) one gets a
set of algebraic equations for the expansion constants $g_{1,2}$:
\begin{eqnarray}
 (\mathbf{p} \mathbf{v}_{11} -\varepsilon )g_{1}(\mathbf{p})+
 \left(\mathbf{p }\mathbf{v}_{12}\right)g_{2}(\mathbf{p}) =0; \nonumber \\
 \left(\mathbf{p }\mathbf{v}_{21}\right)g_{1}(\mathbf{p})  +
 (\mathbf{p} \mathbf{v}_{22} -\varepsilon )g_{2}(\mathbf{p}) =0;
 \label{gEquation}
\end{eqnarray}
where the quasi-particle energy $\varepsilon$  is measured from the degeneration energy,
$\varepsilon_{1}(0) = \varepsilon_{2}(0)=0$, and
the matrix elements of the velocity operator are
\begin{eqnarray}
\mathbf{v}_{\alpha \alpha^{\prime}}= \int u_{\alpha,0}^{\star}(\mathbf{r})
\hat{\mathbf{v}}u_{\alpha^{\prime},0}(\mathbf{r})d\mathbf{r}
 \label{velocity}
\end{eqnarray}

Equating the determinant of  Eq.(\ref{gEquation}) to zero one gets the conventional dispersion
law of quasi-particles near the degeneration point:
\begin{eqnarray}
\varepsilon_{\pm}(\mathbf{p})=\frac{\mathbf{pv_{+}}\pm
\sqrt{(\mathbf{pv_{-}})^2 + 4 |\mathbf{pv}_{12}|^2}}{2}
 \label{ConventionalDispersion}
\end{eqnarray}
where  $v_{\pm}
=\mathbf{v}_{11} \pm \mathbf{v}_{22}$. From here it follows that  the  dispersion  law of  quasi-particles in
the vicinity of  the band intersection is of the  graphene-type (see, e.g., review papers \cite{Neto,Sarma})
\begin{eqnarray}
\varepsilon_{\pm}(\mathbf{p})=\pm v \sqrt{p_x^2 + p_y^2}= \pm vp
 \label{GrapheneDispersionLaw}
\end{eqnarray}
if the lattice symmetry imposes the following conditions  on the velocity matrix elements at
the degeneration  point $p=0$:
\begin{eqnarray}
\mathbf{v}_{11}(0)=\mathbf{v}_{22}(0)=0, \hspace{0.2cm} |\mathbf{v}_{12}(0)| =v, \nonumber
\\
 v_{12}^{(y)}(0) =\pm i v_{12}^{(x)}(0) \hspace{1.5cm}
 \label{VelocityMatixElements}
\end{eqnarray}
where $v=v_F \approx 1 \times 10^6$ m/s for graphene.

Inserting the values of the velocity matrix elements Eq.(\ref{VelocityMatixElements})
in Eq.(\ref{gEquation}), solving the latter equation and using Eq.(\ref{PerturbExpasion}) one finds the graphene Bloch functions
\begin{eqnarray}
\varphi_{\alpha, \mathbf{p}}^{(gr)}(r)=e^{i\mathbf{pr}}\left[ u_{1,0}(\mathbf{r})+e^{-\alpha \theta} u_{2,0}(\mathbf{r}\right]/2\sqrt{v}
 \label{GrafWF}
\end{eqnarray}
where $\theta=\arctan(p_y/p_x)$ and the energy band number is $\alpha=\pm$.


  Introducing the envelope functions
\begin{eqnarray}
  \Phi_{1,2}(\mathbf{r})   = \int g_{1,2} (\mathbf{p})\exp\{i \frac{p \mathbf{r}}{\hbar}\}
  \frac{d\mathbf{p}}{(2 \pi \hbar)^2}
 \label{DiracSimple}
\end{eqnarray}
and using Eqs.(\ref{gEquation})  one finds the following equation:
\begin{eqnarray}
 \left(-i\hbar\mathbf{v}_{11}\partial_\mathbf{r} +V(\mathbf{r})-\varepsilon \right) \Phi_{1}(\mathbf{r})
-i\hbar\mathbf{v}_{12}\partial_\mathbf{r} \Phi_{2}(\mathbf{r})  =0; \;\nonumber \\
-i\hbar\mathbf{v}_{21}\partial_\mathbf{r} \Phi_{1}(\mathbf{r})   +
 \left(-i\hbar\mathbf{v}_{22}\partial_\mathbf{r}+V(\mathbf{r}) -\varepsilon \right) \Phi_{2}(\mathbf{r})  =0\;\;
 \label{WeylDiffEquation}
\end{eqnarray}
Here we added an external potential $V(\mathbf{r})$ which smoothly changes at the atomic scale  (it can be rigorously proved
 as it is shown  in Ref.\cite{scattprobl}).

 Eq.(\ref{WeylDiffEquation}) transforms into Weyl equation
\begin{eqnarray}
\sigma_0 \varepsilon \Phi_W + \sigma_x \partial_x \Phi_W + \sigma_y \partial_y \Phi_W + \sigma_z \partial_z \Phi_W=0
 \label{VelocityMatixElementsWeyl}
\end{eqnarray}
where $\sigma_0$ is the unity matrix and $\sigma_x,\sigma_y,\sigma_z$ are the Pauli matrices
if two energy bands of a 3D semi-metal  are degenerated in the vicinity of the Fermi energy and
the lattice symmetry imposes  the following conditions on the velocity  matrix elements:
\begin{eqnarray}
\mathbf{v}_{11}(0)=\mathbf{v}_{22}(0)=0, \hspace{0.1cm} |\mathbf{v}_{12}(0)| =v, \hspace{0.1cm}
\mathbf{v}_{12}=v(1,-i,i)
 \label{VelocityMatixElementsWeyl}
\end{eqnarray}

Differential  equations  Eq.(\ref{WeylDiffEquation}) and Eq.(\ref{VelocityMatixElementsWeyl}) describe  dynamics of various Weil semi-metals in accordance
with their symmetry that determine the velocity  matrix elements, Eq.(\ref{velocity}).

 Choosing  the graphene symmetry (for which
the matrix elements are given by Eq.(\ref{VelocityMatixElements}) one obtains the conventional Dirac equation \cite{Neto,Sarma}:
\begin{eqnarray}
\left(%
\begin{array}{cc}
V(\mathbf{r}) -\varepsilon& \hbar v(-i\partial_x + \partial_y) \\
\hbar v(-i \partial_x-\partial_y)  &V(\mathbf{r})-\varepsilon\\
\end{array}%
\right)\left(%
\begin{array}{c}
  \Phi_1 \\
  \Phi_2\\
\end{array}%
\right) =0
 \label{DiracEquation}
\end{eqnarray}

\section{Scattering of quasi-particles in graphene by impurity. \label{SecImpScatt}}

Here we consider  the elastic scattering of quasi-particles    by an impurity in  graphene. After solving the Schrödinger equation
for a quasi-particle in the periodic crystal potential with an impurity by the $kp$-method, we derive the  Dirac equation with an effective
scattering potential for the envelope function. The scattering amplitude and Green's functions for the Schr\"{o}dinger and Dirac equations
are also found.

Elastic  scattering of a free quasi-particle by an impurity in the periodic lattice is described by the following Schrödinger equation:
\begin{eqnarray}
\left(\hat{H}_0 +V_i(\mathbf{r}) \right)\Psi(\mathbf{r})= \varepsilon \Psi(\mathbf{r})
 \label{SchroedingerImp}
\end{eqnarray}
where $\hat{H}_0$ is the quasi-particle Hamiltonian for the pure crystal, Eq.(\ref{Schroedinger0}), and $V_i(\mathbf{r})$ is the impurity potential.

Using the  Green's function approach one presents the wave function of the scattered quasi-particle as follows:
\begin{eqnarray}
\Psi(\mathbf{r})= \varphi_{\alpha,\mathbf{p}}^{(in)}(\mathbf{r}) +
\int G(\mathbf{r},\mathbf{r}^{\prime})V_i(\mathbf{r^{\prime}}) \Psi(\mathbf{r}^{\prime})d\mathbf{r}^{\prime}
 \label{WFGreen}
\end{eqnarray}
where $\varphi_{\alpha,\mathbf{p}}^{(in)}(\mathbf{r})$ is the incident "graphene" Bloch function, Eq.(\ref{GrafWF}),
 and $ G(\mathbf{r},\mathbf{r}^{\prime})$
is Green's function satisfying the equation
\begin{eqnarray}
\left(\hat{H}_0 -\varepsilon) \right)G(\mathbf{r},\mathbf{r}^{\prime}))= - \delta(\mathbf{r}-r^{\prime});
 \label{GreenEquation}
\end{eqnarray}

Expanding $G(\mathbf{r},\mathbf{r}^{\prime}))$ in the series of Bloch functions $\varphi_{s,\mathbf{p}}(\mathbf{r})$
one finds Green's function as follows:
\begin{eqnarray}
G(\mathbf{r},\mathbf{r}{\prime}))=G_{\alpha}(\mathbf{r},\mathbf{r}^{\prime}))+G_{s\neq\alpha}(\mathbf{r},\mathbf{r}^{\prime}))
\label{GreenExpandSum}
\end{eqnarray}
where
\begin{eqnarray}
G_{\alpha}(\mathbf{r},\mathbf{r}^{\prime}))=
\sum_{\alpha=1}^{2}\int \frac{d \mathbf{p}}{(2 \pi \hbar)^2}
\frac{\varphi_{\alpha,\mathbf{p}}^{\star}(\mathbf{r}^{\prime})
\varphi_{\alpha,\mathbf{p}}(\mathbf{r})}{\varepsilon -\varepsilon_{\alpha}
(\mathbf{p})+i0} \;\;\;
 \label{WFGreenExpand1}
\end{eqnarray}
is  the "graphene" Green function in which the graphene dispersion law and  the Bloch functions $\varphi_{\alpha,\mathbf{p}}(\mathbf{r}) $   are defined in Eq.(\ref{GrapheneDispersionLaw}) and  Eq.(\ref{GrafWF}), respectively,
while
\begin{eqnarray}
G_{s \neq\alpha}=\sum_{s \neq \alpha}\int \frac{d \mathbf{p}}{(2 \pi \hbar)^2}
\frac{
\varphi_{s,\mathbf{p}}^{\star}(\mathbf{r}^{\prime})\varphi_{s,\mathbf{p}}
(\mathbf{r})}{\varepsilon -\varepsilon_{s}(\mathbf{p})+i0} \Big \}
 \label{WFGreenExpand2}
\end{eqnarray}
is the Green function of virtual states in which  the Bolch functions, Eq.(\ref{Bloch}),
are proper functions of quasi-particle energies $\varepsilon_s(\mathbf{p})$ belonging to
 other   bands, $s\neq\alpha $.

\textbf{a}) \textbf{Calculations of  "graphene" Green's function.}

Using Eqs.(\ref{WFGreenExpand1},\ref{GrafWF}) one presents $G_{\alpha}(\mathbf{r},\mathbf{r}^{\prime}))$ as
follows:
\begin{eqnarray}
G_{\alpha}(\mathbf{r},\mathbf{r}^{\prime}))=\nonumber \\
\frac{1}{2} \Big[u_{1,0}^{\star}(\mathbf{r}^{\prime})u_{1,0}(\mathbf{r})
+ u_{2,0}^{\star}(\mathbf{r}^{\prime})u_{2,0}(\mathbf{r})\Big]
I_1 \hspace{1cm}  \nonumber \\
+\frac{(-1)^{\alpha}}{2}\Big[u_{1,0}^{\star}(\mathbf{r}^{\prime})u_{2,0}(\mathbf{r})I_2^{(+)}
+u_{2,0}^{\star}(\mathbf{r}^{\prime})u_{1,0}(\mathbf{r})I_2^{(-)}\Big] \;\,
 \label{GreenTransform}
\end{eqnarray}
where
\begin{eqnarray}
I_{1}(\mathbf{r}-\mathbf{r}^{\prime}))=
 \int \frac{d \mathbf{p}}{(2 \pi \hbar)^2}
\frac{ e^{i \mathbf{p}(\mathbf{r}-\mathbf{r}^{\prime})/\hbar }
}{\varepsilon -\varepsilon_{\alpha}
(\mathbf{p})+i0}\nonumber \\
I_{2}^{(\pm)}(\mathbf{r}-\mathbf{r}^{\prime}))=\int \frac{d \mathbf{p}}{(2 \pi \hbar)^2}
\frac{ e^{ \mathbf{p}(\mathbf{r}-\mathbf{r}^{\prime})/\hbar}e^{\pm i \theta_p}}{\varepsilon -\varepsilon_{\alpha}
(\mathbf{p})+i0}
 \label{I}
\end{eqnarray}


Performing integrations (see Appendix \ref{AppGraphContourCalc}) one finds
\begin{eqnarray}
I_{1}=I_{2}^{(\pm)} =-\frac{e^{i \pi/4} }{\hbar v}\sqrt{\frac{p_\varepsilon}{2 \pi \hbar }}
\frac{e^{i p_{\varepsilon}R/\hbar}}{\sqrt{R}}
-\frac{ \pi e^{i\pi/4 }}{\varepsilon R^2 }
 \label{I2final}
\end{eqnarray}
where $p_\varepsilon=\varepsilon /v$  is the quasi-particle momentum.

\textbf{b}) \textbf{Calculations of  Green's function for virtual states.}

Here we calculate the  part of Green's function determined
by virtual states, Eq.(\ref{WFGreenExpand2}):

\begin{eqnarray}
G_{s\neq \alpha}(\mathbf{r}^\prime,\mathbf{r}) =\sum_{s \neq \alpha}
\int \frac{u_{s;\mathbf{p}}^{\star}(\mathbf{r}^\prime)u_{s;\mathbf{p}}(\mathbf{r})e^{i p (\mathbf{r-r}\prime)}}{\varepsilon- \varepsilon_s(\mathbf{p})}\frac{d \mathbf{p}}{(2\pi \hbar)^2}
 \label{VirtI}
\end{eqnarray}


In  the polar coordinates the integral in Eq.(\ref{VirtI}) reads
\begin{eqnarray}
G_{s\neq \alpha}(\mathbf{r}^\prime,\mathbf{r}) \hspace{4.5cm} \nonumber \\ = \sum_{s \neq \alpha} \int_0^{\infty}\frac{d p \, p}{(2\pi \hbar)^2} \int_0^{2 \pi}d \varphi \frac{U_s(p, \varphi) e^{i p R \cos{\varphi}}}{\varepsilon- \varepsilon_s(p, \varphi)}
 \label{VirtIpolar}
\end{eqnarray}
where $$U_s(\mathbf{p})=u_{s;\mathbf{p}}^{\star}(\mathbf{r}^\prime)u_{s;\mathbf{p}}(\mathbf{r}); \hspace{0.5cm}R=|\mathbf{r'-r}|$$
with the momenta taken in the polar coordinates.

\begin{figure}
\centerline{\includegraphics[width=6.0cm]{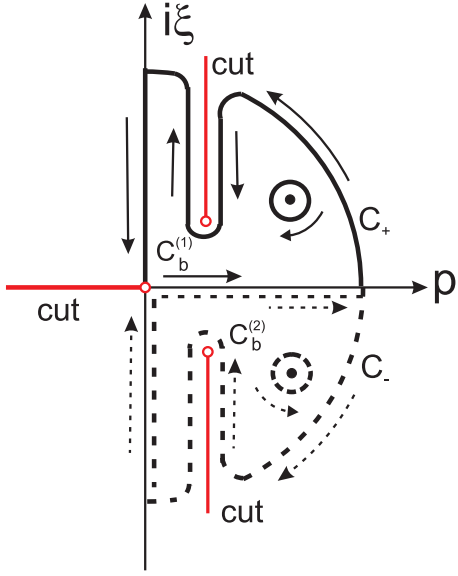}}
\caption{Closed contours of integration $C_+$ and  $C_-$ for calculations of the first and second integrals
in Eq.(\ref{VirtIpolar}) are shown with solid and dotted lines, respectively; branching points
  are shown with small circles
 at the beginnings of branch cuts; poles of the integrands are shown with black dots.}
\label{FigContourVirtual}
\end{figure}

At  $Rp/\hbar \gg1$ one may use the fastest descent method for calculations of the integral with respect to $\varphi$ and find
\begin{eqnarray}
G_{s\neq \alpha}(\mathbf{r}^\prime,\mathbf{r}) = \sqrt{\frac{2 \pi \hbar}{R}} \sum_{s \neq \alpha}
 \int_0^{\infty} \frac{d p \sqrt{p}}{(2\pi \hbar)^2}
\nonumber \\
\Big\{
\frac{U_s(p,0)e^{-i \pi/4}}{\varepsilon- \varepsilon_s(p,0)}
 e^{i p R}
+ \frac{U_s(p,\pi)e^{i \pi/4}}{\varepsilon- \varepsilon_s(p,\pi)} e^{-i p R}  \Big\}
 \label{VirtIpolarFD}
\end{eqnarray}
For calculations of the above integrals it is convenient to choose the integration contours in the complex plane
  shown in Fig.\ref{FigContourVirtual}.

In the general case the dispersion equations $\varepsilon_s(p, \varphi)$  considered as  functions of
the complex variable $z=p+i\xi$ have branching points, their
 characteristic distances   from the real axis being of the order of $\hbar/ a$ (here $a$ is the
 atomic spacing). In Fig.\ref{FigContourVirtual}, they are schematically shown with small circles
 at the beginnings of branch cuts; as the energy $\varepsilon$ is out of the energy band under consideration
 $s \neq \alpha$  the poles (which are shown with black dots) are  in the complex planes with  $|\xi| \gtrsim \Delta/v$
 where $\Delta$ is the characteristic width of  energy gaps.

Performing the contour integrations in the complex plane (see Appendix \ref{AppGVirtual} ) one finds Green's function
for virtual states:
 \begin{eqnarray}
G_{s\neq \alpha}(\mathbf{r}^\prime,\mathbf{r})\sim \frac{e^{-R/a}}{\hbar v \sqrt{2 \pi  a R}}+\frac{1}{R^2\Delta}
 \label{VirtGreenFinal}
\end{eqnarray}

Finally, according to Eqs.(\ref{GreenExpandSum},\ref{GreenTransform},\ref{I2final},\ref{VirtGreenFinal})
the total Green function for the electron reads

\begin{eqnarray}
G_{e}(\mathbf{r},\mathbf{r}^{\prime}))=
-\frac{e^{i \pi/4} }{2 \hbar v}\sqrt{\frac{p_\varepsilon}{2 \pi \hbar }}
\frac{e^{i R p_\varepsilon}}{\sqrt{R}}  \nonumber \\
\times \sum_{\alpha,\beta =1}^2 u_{\alpha,0}^{\star}(\mathbf{r}^{\prime})
 u_{\beta,0}(\mathbf{r}^{\prime})
+  \mathcal{O}\left(\frac{1}{\Delta  R^2 } , \frac{e^{-R/a}}{v \sqrt{a R}} \right)
 \label{GreenFinal1}
\end{eqnarray}

Inserting Eq.(\ref{GreenFinal1}) into Eq.(\ref{WFGreen}) one readily  finds the integral equation for the wave function of the electron scattered by the impurity in graphene:
  \begin{eqnarray}
\Psi(\mathbf{r})= \varphi_{\alpha,\mathbf{p}}^{(in)}(\mathbf{r}) -\frac{(2 \pi)^{3/2}
\sqrt{\hbar p_\varepsilon}}{v}e^{i\pi/4}\times \hspace{1.5cm} \nonumber \\
\sum_{\alpha,\beta=1}^2u_{\alpha;0}(\mathbf{r})
\int u_{\beta;0}(\mathbf{r^\prime}) V_i(\mathbf{r^{\prime}}) \Psi(\mathbf{r}^{\prime})\frac{e^{i p_\varepsilon|\mathbf{r-r^\prime}|}}{\sqrt{|\mathbf{r}-\mathbf{r}^\prime|}}\frac{d\mathbf{r}^{\prime}}{(2\pi\hbar)^2} \nonumber \\
+  \mathcal{O}\left(\frac{a^2}{R^2 } , \frac{e^{-R/a}}{ \sqrt{ R/a}} \right)  \hspace{4.7cm}
 \label{WFGreenAsymp}
\end{eqnarray}
This equation can be easily solved in Born's or semiclassical approximations that gives the explicit expression for the
wave function of the electron scattered  by the impurity.

As one sees from Eq.(\ref{WFGreenAsymp}), in the vicinity of the impurity the  wave function of the quasi-particle   scattered  by the impurity is a superposition of the virtual states belonging to all  available energy bands that fast decays as the distance from the impurity increases.

In the next section, using the \textbf{k-p}-method we derive the Dirac equation for
quasi-particles in graphene with an impurity. As is shown there solution of   this equation in Born's approximation allows
to  present the envelope function and the scattering amplitude in an explicit form.

\textbf{ Envelope function and  scattering amplitude for graphene with an impurity.}

First we derive the Dirac equation for quasi-particles in graphene with an impurity using  the \textbf{k-p} method.
For this purpose we write the Schr\"{o}dinger equation considering the term with the impurity potential as a known
function in the right-hand side of it:
\begin{eqnarray}
\left[-\frac{\hbar^2}{2 m}\frac{\partial^2}{\partial\mathbf{ r}^2} +U(\mathbf{r})- \varepsilon \right]
\Psi(\mathbf{r})=-V_i(\mathbf{r})\Psi
(\mathbf{r})
 \label{SchroedingerImp1}
\end{eqnarray}

Expanding $\Psi$ in the left-hand side of the above equation in the series of $\chi$  (see Eq.(\ref{chi}))
\begin{eqnarray}
\Psi=\sum_{\alpha =1}^2\int g_\alpha(\mathbf{p})\chi_{\alpha, \mathbf{p}}(\mathbf{r})
\frac{d \mathbf{p}}{(2 \pi \hbar)^2}
 \label{PsiExpand}
\end{eqnarray}
and using Eq.(\ref{VelocityMatixElements}) one finds the Schr\"{o}dinger equation in the $\mathbf{p}$-representation:
\begin{eqnarray}
 -\varepsilon g_\alpha(\mathbf{p}) + \sum_{\alpha^\prime =1}^2(\mathbf{p}\cdot \mathbf{v}_{\alpha,\alpha^\prime})
 g_{\alpha^\prime}(\mathbf{p}) \nonumber \\
 =-\int \chi^\star_{\alpha,\mathbf{p}}(\mathbf{r})V_i(\mathbf{r})\Psi(\mathbf{r})d\mathbf{r}
 \label{gEquation1}
\end{eqnarray}
In the above equation,    contributions of the virtual states are neglected (see the previous section).

The  envelope functions are given by Eq.(\ref{{DiracSimple}})
and hence, according to Eq.(\ref{PsiExpand}), they  are related to  the wave function of the Schr\"{o}dinger equation, Eq.(\ref{SchroedingerImp}),
by the following relation:
\begin{eqnarray}
\Psi = \sum_{\alpha=1}^2 u_{\alpha,0}(\mathbf{r})\Phi_{\alpha}(\mathbf{r})
 \label{PsiPhi}
\end{eqnarray}

After multiplying the both sides of Eq.(\ref{gEquation1})  by $\exp\{i \mathbf{p r}/\hbar)\}$ and integrating with respect to $\mathbf{p}$ one finds the following equation
for the envelope function:
 \begin{eqnarray}
\left\{%
\begin{array}{cc}
\varepsilon \Phi_1 +\hbar  v(i \partial_x-\partial_y )
\Phi_2
 =u_{1,0}(\mathbf{r})V_i(\mathbf{r})\Psi(\mathbf{r}) \\
\hbar  v(i  \partial_x+\partial_y )
\Phi_1 +\varepsilon  \Phi_2
=u_{2,0}(\mathbf{r})V_i(\mathbf{r})\Psi(\mathbf{r})
\end{array}%
\right.
 \label{DiracImpquation}
\end{eqnarray}

Treating the right-hand side as a known function one  finds the following solution of this Dirac equation:
\begin{eqnarray}
 \Phi_1(\mathbf{r})=\Phi_1^{(in)} -\int V_i(\mathbf{r}^\prime)\Psi(\mathbf{r}^\prime)\sum_{\alpha=1}^2
 u^{\star}_{\alpha,0}(\mathbf{r}^\prime)A_{\alpha}(\mathbf{r}^\prime,\mathbf{r})
 d \mathbf{r}^\prime \nonumber \\
 \Phi_2(\mathbf{r})= \Phi_2^{(in)} -\int V_i(\mathbf{r}^\prime)\Psi(\mathbf{r}^\prime)\sum_{\alpha=1}^2
 u^{\star}_{\alpha,0}(\mathbf{r}^\prime)B_{\alpha}(\mathbf{r}^\prime,\mathbf{r})
 d \mathbf{r}^\prime \;
 \label{DirImpSolution}
\end{eqnarray}
where $\Phi_{1,2}^{(in)}$ are the envelope functions of the incoming quasi-particle (which are solutions of the
above homogeneous Dirac equation) while
functions $A_{\alpha}$ and $B_{\alpha}$ are integrals with respect to the momentum $\mathbf{p}$:
\begin{eqnarray}
A_1=B_1=\varepsilon \int_{-\infty}^{\infty}\frac{e^{ip(\mathbf{r-r}^\prime)}}{(v p)^2 -\varepsilon}
\frac{d \mathbf{p}}{(2\pi \hbar)^2} \nonumber \\
A_2 = \int_{-\infty}^{\infty}\frac{v(p_x+ip_y)e^{ip(\mathbf{r-r}^\prime)}}{(v p)^2 -\varepsilon}
\frac{d \mathbf{p}}{(2\pi \hbar)^2}\nonumber \\
B_2 = \int_{-\infty}^{\infty}\frac{v(p_x-ip_y)e^{ip(\mathbf{r-r}^\prime)}}{(v p)^2 -\varepsilon}
\frac{d \mathbf{p}}{(2\pi \hbar)^2}
 \label{ABdefinition}
\end{eqnarray}

Performing integrations analogous to those made in Appendix\ref{AppGraphContourCalc}
one finds:edge
\begin{eqnarray}
 A_1=A_2=B_1=B_2=e^{i\pi/4} \frac{(2\pi)^{3/2}}{4 (2\pi \hbar)^2}\frac{p_\varepsilon}{v}
 \frac{e^{i p_\varepsilon R/\hbar}}{\sqrt{p_\varepsilon R/\hbar}}; \nonumber \\
R=|\mathbf{r} - \mathbf{r}^\prime|; \hspace{5.7cm}
 \label{AB}
\end{eqnarray}

Inserting Eq.(\ref{AB}) into Eq.(\ref{DirImpSolution}) one finds  the set of integral equations for the envelope functions
 of the graphene with an impurity as follows:
\begin{eqnarray}
\left(\begin{array}{c}
  \Phi_1 \\
  \Phi_2\\
\end{array}%
\right)=
\left(\begin{array}{c}
  1 \\
e^{i \varepsilon}\\
\end{array}%
\right)       e^{i\mathbf{\mathbf{r} k}_\varepsilon }
 -\left(\begin{array}{c}
  +1 \\
-1\\
\end{array}%
\right)\int V_i(\mathbf{r}^\prime)  \nonumber \\
\times \sum_{\beta=1}^2
u^{\star}_{\beta, 0} (\mathbf{r}^\prime) \Phi_\beta (\mathbf{r}^\prime)
\sum_{\alpha=1}^2u^{\star}_{\alpha, 0} (\mathbf{r}^\prime)
A_1(\mathbf{r}^\prime , \mathbf{r})d\mathbf{ r}
\label{DiracEquationImp}
\end{eqnarray}
where  $\varphi=\arctan{p_y/p_x}$ and  for the sake of definiteness, the scattering of an electron is considered.  While writing this equation Eq.(\ref{PsiPhi}) was used.

In the Born approximation  the second term in the right-hand side of Eq.(\ref{DiracEquationImp} is considered as a perturbation and
 at large distances from the impurity  one
finds the envelope function of the electron scattered by the impurity as follows:
\begin{eqnarray}
\left(\begin{array}{c}
  \Phi_1 \\
  \Phi_2\\
\end{array}%
\right)=
\left(\begin{array}{c}
  1 \\
e^{i \varphi}\\
\end{array}%
\right)       e^{i\mathbf{\mathbf{r} k}_\varepsilon }
 -\left(\begin{array}{c}
  +1 \\
-1\\
\end{array}%
\right)f(\theta) \frac{e^{ik_{\varepsilon}R_0}}{\sqrt{R_0}}
\label{DiracBorn}
\end{eqnarray}
where the scattering amplitude is
\begin{eqnarray}edge
f(\theta)=-\frac{\pi^{3/2 }}{\sqrt{2}} e^{i \pi/4}
 \sqrt{\frac{\hbar v}{\varepsilon}}\times \hspace{4cm} \nonumber \\
 \frac{\varepsilon}{v^2}
\int\frac{ d \mathbf{r}^\prime   }{(2\pi \hbar)^2}e^{-i \mathbf{qr}^\prime} V_i(\mathbf{r}^\prime)
 \sum_{\alpha,\beta=1}^2u_{\alpha,0}^\star(\mathbf{r}^\prime)
u_{\beta,0}(\mathbf{r}^\prime) e^{i(\beta-1)\varphi}\;
 \label{ScatterinAmpl}
\end{eqnarray}
While writing the above equation we chose the coordinate origin at the scattering center
and introduced the radius vector $\mathbf{R}_0$  from the origin to the observation point,
 a unity vector along it being denoted  by $\mathbf{n}^\prime$. Therefore, in this coordinates
vector $\mathbf{R}$ (see Eq.(\ref{AB})) reads $\mathbf{R}=\mathbf{R}_0 -\mathbf{r}^\prime$.
At large distances from the center, $R_0 \gg |\mathbf{r}^\prime|$, one has
$R \approx R_0-\mathbf{k}^\prime\mathbf{n}^\prime$. Vector $\mathbf{q}=\mathbf{k}^\prime-\mathbf{k}$
where $\mathbf{k}^\prime=k \mathbf{n}^\prime$ is the wave vector of the quasi-particle after scattering;
 $$q =2 k \sin{\theta/2},$$ $\theta$
being the angle between $\mathbf{k}$ and $\mathbf{k}^\prime$, i.e. the scattering angle.

As one sees the  envelope function, Eq.(\ref{DiracBorn}),
 and Dirac equation equation for it, Eq.(\ref{DiracEquation}), are tightly coupled withedge
 the wave function, Eq.(\ref{WFGreenAsymp}), and Schr\"{o}dinger equation, Eq.(\ref{SchroedingerImp}) via
 the function-envelope function relation  Eq.(\ref{PsiPhi}).   Below we present  a Green's function equation
for the Dirac equation which is closely associated with Green's function of the Schr\"{o}dinger equation.

\textbf{Green's function for the Dirac equation}.
 Green's functions  are  convenient  tools for investigations of properties of various  systems  and
 it may be desirable to have  an  equation for Green's function for the Dirac equation, Eq.(\ref{DiracEquation}),
  closely related to the Schr\"{o}dinger equation, Eq.(\ref{Schroedinger0}), and the  corresponding
  Green's function equation, Eq.(\ref{GreenEquation}).

Using Eq.(\ref{GreenEquation}) for  Green's function $G(\mathbf{r},\mathbf{r}^\prime)$ of the Schr\"{o}dinger equation,
Eq.(\ref{Schroedinger0}) and repeating the reasoning for  derivation  of Eq.(\ref{DiracImpquation}) from Eq.(\ref{SchroedingerImp1})
one finds the equation for  Green's function of the Dirac equation as follows:
\begin{eqnarray}
\left(%
\begin{array}{cc}
-\varepsilon   &\hbar v (-i\partial_x + \partial_y )\\
\hbar v(-i \partial_x- \partial_y ) &-\varepsilon\\
\end{array}%
\right)\left(%
\begin{array}{c}
 G^{(D)}_1(\mathbf{r},\mathbf{r}^\prime) \\
  G^{(D)}_2(\mathbf{r},\mathbf{r}^\prime)\\
\end{array}%
\right)\nonumber \\
 = -\left(%
\begin{array}{c}
u_{1,0}(\mathbf{r}) \\
u_{2,0}(\mathbf{r})\\
\end{array}%
\right)\delta(\mathbf{r}-\mathbf{r}^\prime)\hspace{4cm}
 \label{GreenDiracEquation}
\end{eqnarray}

In Eq.( \ref{GreenDiracEquation}), expanding  in the series of the proper function of the Dirac equation one finds
that Green's function reads as follows:edge
\begin{eqnarray}
\left(%
\begin{array}{c}
 G^{(D)}_1(\mathbf{r},\mathbf{r}^\prime) \\
  G^{(D)}_2(\mathbf{r},\mathbf{r}^\prime)\\
\end{array}%
\right)
 = \sum_{\alpha=1}^2\int \frac{d \mathbf{p}}{(2 \pi \hbar)^2} \hspace{2.5cm}  \nonumber \\
 \times \frac{u_{1,0}(\mathbf{r})+(-1)^{\alpha}e^{i\theta}u_{2,0\mathbf{r}}}{\varepsilon -\varepsilon_\alpha(\mathbf{p})}
 \left(%
\begin{array}{c}
1 \\
(-1)^\alpha e^{-i\theta}\\
\end{array}%
\right)e^{\mathbf{p}(\mathbf{r}-\mathbf{r}^\prime)/\hbar}
 \label{GreenDirac}
\end{eqnarray}
where $\varepsilon_\alpha(\mathbf{p})=(-1)^{\alpha}vp$ and $\theta=\arctan(p_x/p_y)$.

\section{Derivation of  boundary conditions for  Dirac equation. \label{BoundaryConditions}}
edge
Dynamics of quasiparticles in graphene that occupies  the upper half plane $y \geq 0$ is described by
 Schr\"odinger equation
\begin{eqnarray}
\left[-\frac{\hbar^2}{2 m}\frac{\partial^2}{\partial\mathbf{ r}^2} +U(\mathbf{r})\right] \Psi(\mathbf{r})=\varepsilon
 \Psi(\mathbf{r})
 \label{SchroedingerSharp}
\end{eqnarray}
with the boundary condition
\begin{eqnarray}
\Psi(\mathbf{r})\Big|_{y=0}=0
 \label{BoundaryCondition1}
\end{eqnarray}
where $U(\mathbf{r})=U(\mathbf{r}+\mathbf{a})$ is the lattice periodic  potential

To solve  the  problem of reflection by the sharp edge at $y=0$, we use Green's function
for  Schr\"odinger equation Eq.(\ref{SchroedingerSharp}):
\begin{eqnarray}
\left(-\frac{\hbar^2}{2 m}\frac{\partial^2}{\partial\mathbf{ r}^2} +U(\mathbf{r})
-\varepsilon\right)G(\mathbf{r},\mathbf{r}^{\prime})=\delta(\mathbf{r}-\mathbf{r}^{\prime})
 \label{GreenFunctionEq}
\end{eqnarray}
in which the  lattice potential $U(\mathbf{r})$ covers the whole plane $(x,y)$.

Using Eqs.(\ref{GreenEquation},\ref{SchroedingerSharp}) and taking into account the boundary condition Eq.(\ref{BoundaryCondition1})
one finds
\begin{eqnarray}
\Psi(\mathbf{r})=\frac{\chi_{\alpha,\mathbf{p}}^{(in)}({\bf r})}{\sqrt{v_{y,\alpha}}}   \hspace{3cm}       \nonumber \\
+\frac{\hbar^2}{2 m}
\int_{-\infty}^{+\infty}G(x^{\prime},-0;\mathbf{r})
\frac{\partial \Psi(\mathbf{r}^{\prime})}{\partial y^{\prime}}\Big|_{y^{\prime}=-0}d x^{\prime}
\label{PsiG}
\end{eqnarray}edge
Here $\chi_{\alpha,\mathbf{p}}^{(in)}({\bf r})$ is the graphene Kohn-Lattinger function Eq.(\ref{chi}) incident to the graphene edge from the
infinity $y \rightarrow \-\infty$ and $v_{y,\alpha}= \partial \varepsilon_\alpha^{(gr)}(\mathbf{p}) /\partial y$ is the velocity
$y$-projection that normalizes the incident function to the flux unity while $\varepsilon_\alpha^{(gr)}(\mathbf{p})= \pm v p$ is the graphene dispersion; in order to define $\Psi(\mathbf{r})$
on the whole half-plane $y \geq 0$  the boundary contour is shifted to $y=-0
\equiv 0-\delta^{\prime}, \;\delta^{\prime} \rightarrow 0$ (see Ref.\cite{Morse}).

 Expanding
$G(\mathbf{r},\mathbf{r}^{\prime})$  in the series of Bloch wave functions   and using Eq.(\ref{GreenFunctionEq}) one  finds
\begin{eqnarray}
G(\mathbf{r},\mathbf{r}^{\prime})= \sum_{\alpha=1,2}\int\frac{\chi_{\alpha,\mathbf{p}}^{\star}(\mathbf{r})
\chi_{\alpha,\mathbf{p}}(\mathbf{r}^{\prime})}{\varepsilon-\varepsilon_\alpha^{(gr)}(\mathbf{p})+i \delta}d
\mathbf{p} \nonumber \\
 +\sum_{s \neq 1,2}\int\frac{\varphi_{s,\mathbf{p}}^{\star}(\mathbf{r})
\varphi_{s,\mathbf{p}}(\mathbf{r}^{\prime})}{\varepsilon-\varepsilon_s(\mathbf{p})+i \delta}d
\mathbf{p}
 \label{GreenFunctionExpanded}
\end{eqnarray}
where summation goes over all energy bands  and
 $\delta \rightarrow +0$

Inserting Eq.(\ref{GreenFunctionExpanded}) into Eq.(\ref{PsiG}) one finds
the wave function  on the right half-plane $x \geq 0$ as follows:
\begin{eqnarray}
\Psi(\mathbf{r})=   \frac{\chi_{\alpha,\mathbf{p}_0}^{(in)}}{\sqrt{v_{y,\alpha}}} + \hspace{5.2cm}\nonumber \\
 \frac{\hbar^2}{2 m}\int_{-\infty}^{\infty} d\bar{x}
\Psi_{y}^{\prime}(\bar{x},0)e^{ix p_x\hbar}
\Big\{\sum_{\alpha = 1,2} u^{\star}_{\alpha,0}(\bar{x},0)u_{\alpha,0}(\mathbf{r})I_\alpha^{(gr)}
 \nonumber \\
+\sum_{s\neq 1,2}  u^{\star}_{s,\mathbf{p}}(\bar{x},0)u_{s,\mathbf{p}}(\mathbf{r})I_s^{(bnd)} \Big\}\;\;
\label{FunctionPsi}
\end{eqnarray}
where$\Psi_y^{\prime}(\bar{x},-0)=\partial\Psi(\mathbf{r})/\partial y$ at $y=-0$. While writing the above equation
we assumed that  along the edge line $y=0$ the lattice is periodic with the period $a_x$ that is $\Psi(x,0) =\Psi(x+a_x,0)$
and hence the momentum projection $p_x$ conserves; $I_\alpha^{(gr)}$ and $I_s^{(bnd)}$ are one-dimensional integrals defined
below, Eqs.(\ref{ComplexGraphIntegral},\ref{ComplexBandsIntegral})

Differentiating the both sides of Eq.(\ref{FunctionPsi}) with respect to $y$ one obtains
the integral equation for $\Psi_y^{\prime}(\bar{x},-0)$ the solution of which completes the
definition of the  sought  wave function $\Psi(\mathbf{r})$.
Despite   this integral equation can not
be solved in the general case  important properties
of the quasi-particle scattering    by the sharp  sample boundary  may be derived  from  Eq.(\ref{FunctionPsi}).

Indeed, let us consider  one-dimensional integrals with respect to $p_y$ in Eq.(\ref{FunctionPsi}) re-writing them
 in the following forms:
\begin{eqnarray}
I_\alpha^{(gr)}=\int_{-b_y/2}^{b_y/2} \frac{e^{i y p_y/\hbar}}{\varepsilon-\varepsilon_\alpha(p_x,p_y)+i \delta}dp_y
 \label{ComplexGraphIntegral}
\end{eqnarray}
and
\begin{eqnarray}
I_s^{(bnd)}=\int_{-b_y/2}^{b_y/2} \frac{u_{s,p_x,p_y}^{\star}(\bar{x},0)
u_{s,p_x,p_y}(\mathbf{r})e^{iy p_y/\hbar}}{\varepsilon-\varepsilon_s(p_x,p_y)+i \delta}dp_y
 \label{ComplexBandsIntegral}
\end{eqnarray}
Here $b_y$ is the period of the reciprocal lattice in the $y$-direction.

 In the complex plane  the dispersion law   of the degenerated bands of graphene Eq.(\ref{GrapheneDispersionLaw}) considered as a function of the complex variable $z=p_y+i\xi$ (that is $\varepsilon(p_x,z) =+v\sqrt{z^2+p_x^2}$)  has  branch points at $z= \pm i p_x$ and the two branches of this complex function are the two  energy bands   on the real axis $z=p_y$. The dispersion law functions of other energy bands are also multi-valued functions with branch points in the complex plane.

\begin{figure}
\centerline{\includegraphics[width=\columnwidth]{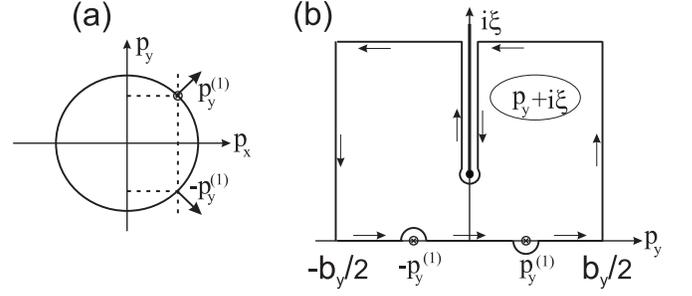}}
\caption{(a) Equal energy contour  $v \sqrt{p_x^2+p_y^2}=\varepsilon$. The thick arrows show the velocity direction
at fixed energy $\varepsilon$  and $p_x$. The incident quasiparticle has conserving projection $p_y=-p_y^{(1)}$ while
the outgoing quasiparticle has $p_y=+p_y^{(1)}$.  (b)
Contour of integration of Eq.(\ref{ComplexGraphIntegral}). Dots on the real
axis $p_y$   show positions of the poles
corresponding to  points with positive and negative velocity $v_y$. Thick vertical line is the
branch line corresponding to the branching point (thick dots), $p_y^{(gr)} =i p_x$,  in the quasi-psrticle
spectrum.}
\label{OneVallyFig}
\end{figure}

Therefore, integral Eq.(\ref{ComplexGraphIntegral}) is a sum of the residues and the integral along  the branch cut in the upper complex half-plane $\xi \geq 0$ inside the contour schematically shown in Fig.\ref{OneVallyFig}. The left and right vertical lines of the contour are separated by the reciprocal  period $b_y$ and hence the integrals along them cancel each other because the integrands are periodic functions of the same  period. The integral along its upper horizontal part exponentially goes to zero  as this contour part goes to $i \infty$.

Below,   for the sake of certainty we consider here  one valley reflection  of an electron, $\alpha=1$.
We also assume that  only one contour $\varepsilon_1(p_x, p_y)=\varepsilon$    exists  at a fixed $p_x$ as shown
in Fig.\ref{OneVallyFig}.

In this case Eq.(\ref{ComplexGraphIntegral})  reads
\begin{eqnarray}
I^{(gr)}_1=\int_{-b_y/2}^{b_y/2} \frac{e^{i y p_y/\hbar}}{[\varepsilon-\sqrt{p_x^2+p_y^2}+i \delta]}dp_y
 \label{ComplexGraphIntegralOneVally}
\end{eqnarray}

The pole of the integrand in Eq.{\ref{ComplexGraphIntegralOneVally}} which
contributes to the integral lies  on the right upper side of the
real axis (see Fig.\ref{OneVallyFig})
 $$p_y=p_1^{(1)}
+i\frac{\delta}{v_x^{(\alpha)}},  \; \delta \rightarrow 0$$ where
its real part is $p_y^{(1)} =\sqrt{(\varepsilon/v)^2-p_x^2}$. One easily sees from the denominator of the integrand
that this pole is  inside the integration contour because the  velocity
 $$v_y =\frac{\partial \varepsilon (p_x,p_y)}{\partial p_y} \Big|_{p_y={p_y^{(1)}}}>0$$
and hence it corresponds to the quasiparticle state reflected back by the boundary.

Taking into account the above-mentioned pole and branch cut one easily  carried out
integration  in  Eq.(\ref{ComplexGraphIntegralOneVally}) (calculations of the integral along the branch cut is
presented in Appendix \ref{contour})
 and finds $I^{(gr)}_1$ as follows:
\begin{eqnarray}
I^{(gr)}_1=-\frac{2 \pi i}{v_y(p_x,p_y^{(1)})}e^{i y p_y^{(1)}/\hbar}
+\frac{2 \hbar i}{y \varepsilon} e^{-y p_x\hbar}
 \label{ComplexGraphIntOneVallyFinal}
\end{eqnarray}

For calculations of  the integral in Eq.(\ref{ComplexBandsIntegral}) one finds the poles from the equation $\varepsilon_s(p_x,p_y)=\varepsilon, \; s \neq \alpha$ where the energy bands $\varepsilon_s(p_x,p_y)$ do not overlap bands $\alpha =1,2$ in which the energy $\varepsilon$. In the general case the difference between those bands $|\varepsilon_\alpha(p_x,p_y)-\varepsilon_s(p_x,p_y^{\prime})| \gtrsim \Delta_{gap}^{(s)} \sim \hbar v/a, s \neq \alpha$ (where $\Delta_{gap}^{(s)}$ is the characteristic value of the energy gap between the energy bands)
and hence poles of the integrand in the upper imaginary plane have large imaginary parts $\xi \sim b_0^{(s)} = \Delta_{gap}^{(s)}/v$. On the other hand the dispersion laws $\varepsilon_s(p_x,z)$ as functions of the complex variable $z=p_y +i \xi$ are also multi-branched, the branching points of which
have also large imaginary parts $\xi \sim b_0^{(s)}$.

Performing  integration in Eq.(\ref{ComplexBandsIntegral}) in much the  same
manner  as above one finds $I_s^{(bnd)}$ as follows (details of the calculations are presented in Ref.\cite{scattprobl}):
\begin{eqnarray}
I_s^{(bnd)} \sim \frac{e^{-y b_0/\hbar}}{v}
 \label{ComplexBandIntOneVallyFinal}
\end{eqnarray}

Using Eqs,(\ref{ComplexGraphIntOneVallyFinal},\ref{ComplexBandIntOneVallyFinal}) together with Eq.(\ref{FunctionPsi})
we found that at distances $y \gg a$ (here $a$ is the characteristic period of the graphene lattice) the  graphene wave fuction
is the difference between the incident and outgoing Bloch functions of the infinite graphene:
\begin{eqnarray}
\Psi_{p_x}(\mathbf{r})= \left( \varphi_{\alpha;p_x, p_y^{(in)}}^{(gr)}(\mathbf{r}) -
\varphi_{\alpha;p_x, p_y{( out)}}^{(gr)}(\mathbf{r})\right) \nonumber \\
+ C \frac{a e^{-y p_x/\hbar}}{y}e^{ix p_x/\hbar}u_{\alpha;0}(\mathbf{r})
 \label{GrapeneWVBoundary}
\end{eqnarray}
where $p_y^{(in)}$ and $p_y^{(out)} = -p_y^{(in)}$ are the $y$-projections of the quasiparticle momentum while $C$  is
a constant $\sim 1$ (details of calculations are given in Ref. \cite{scattprobl}).

From Eq.(\ref{GrapeneWVBoundary}) and Eq.(\ref{PsiPhi}) one easily finds that at the distances from the graphene sharp edge much greater than the atomic spacing, $l\gg a$,  the graphene envelope function $\check{\Phi}(\mathbf{r})$ is the difference between  the incident and outgoing wave functions (which are
 two independent solutions of the Dirac equation
Eq.(\ref{DiracEquation})):
\begin{eqnarray}
\check{\Psi}(\mathbf{r})= e^{ix p_x}\left[e^{iy p_y^{(in)}}\left(%
\begin{array}{c}
  1\\
 e^{i \varphi}\\
\end{array}%
\right)-  e^{-i(y p_y^{(in)})} \left(%
\begin{array}{c}
1 \\
   e^{-i \varphi}\\
\end{array}%
\right)  \right]
 \label{GraphBC}
\end{eqnarray}
where the phase $\varphi =\arctan (p_y^{(in)}/p_x)$

\section{Conclusion.  \label{conclusion}}

In this paper dynamics of quasi-particles in graphene with an impurity and a sharp edge is considered
with the \textbf{kp}-approach.  Dirac equation for graphene and Weyl equation for semi-metals are derived
in section \ref{SecDiracDerivatuion}. For graphene with an impurity, the wave function and its evolution function together with the scattering
amplitude are found in the Born approximation. As an auxiliary tool Green's functions for Schr\"{o}dinger and Dirac equations are also derived.
In the both cases of the impurity and the sharp edge,  the wave functions of the scattered quasi-particles are shown to be superpositions
of virtual states which exponentially decay outward from the scatterer. They are explicitly presented for   distances much greater that the atomic spacing. In the case that the velocity direction of the incident quasi-particle is perpendicular to the edge
the above-mentioned superposition of virtual states decays linear with the distance increase, Eq.(\ref{GrapeneWVBoundary}). At the distances much greater
than the atomic spacing  the graphene envelope function
 is the difference between the incident and outgoing
wave functions  which are two independent solutions
of the Dirac equation for the infinite graphehe, Eq.(\ref{GraphBC}), that is
the boundary condition for Dirac equation.

\textbf{\emph{Acknowledgement}}. This work was supported by Croatian Science Foundation, project IP-2016-06-2289.

\begin{appendix}

\

\section{Calculations of contour integrals for "graphene"  Green's functions. \label{AppGraphContourCalc}}

Inserting the polar coordinates in the integrals  in Eq.(\ref{I}) one finds
\begin{eqnarray}
I_{1}=-\frac{2 \pi}{v}\int_0^{\infty}
\frac{p}{p-p_{\varepsilon}-i 0}J_0(pR)\frac{d\, p}{(2\pi \hbar)^2} \nonumber \\
I_{2}^{(\pm)}=-\frac{2 \pi i}{v}\int_0^{\infty}
\frac{p}{p-p_{\varepsilon}-i 0}J_1(pR)\frac{d\, p}{(2\pi \hbar)^2}
 \label{I2}
\end{eqnarray}
where $p_{\varepsilon}=\varepsilon/v$ and $R=|\mathbf{r}-\mathbf{r}^{\prime}|$ while $J_{0,1}$ are the Bessel functions
of the first kind.
For the sake of certainty, here and below all calculations are done for  electrons
the dispersion law of which is $\varepsilon_+(p)=v p$ (see Eg.(\ref{GrapheneDispersionLaw})).

Asymptotic of the Bessel functions for large arguments are $J_0(pR) = \sqrt{2/(\pi pR)}\cos(pR - \pi/4)$
and $J_1(pR) = \sqrt{2/(\pi Rp)}\sin(pR - \pi/4)$ and hence at $p R \gg 1$, Eq.(\ref{I2}) reads
\begin{eqnarray}
I_{1}(\mathbf{r}-\mathbf{r}^{\prime})=
-\frac{1}{v}\sqrt{\frac{2 \pi}{R}}
\int_0^{\infty} \frac{d\, p}{(2\pi \hbar)^2} \nonumber \\
\times  \frac{\sqrt{p}}{p-p_{\varepsilon}-i 0}
\Big(\exp^{i(pR-\pi/4)} +\exp^{-i(pR-\pi/4)}   \Big)\nonumber \\
I_{2}^{(\pm)}(\mathbf{r}-\mathbf{r}^{\prime})=-\frac{1}{v}\sqrt{\frac{2 \pi}{R}}
\int_0^{\infty} \frac{d\, p}{(2\pi \hbar)^2}\nonumber \\
\times \frac{\sqrt{p}}{p-p_{\varepsilon}-i 0}\Big(\exp^{i(pR-\pi/4)} -\exp^{-i(pR-\pi/4)}   \Big)\;\,
 \label{I21}
\end{eqnarray}

\begin{figure}
\centerline{\includegraphics[width=6.0cm]{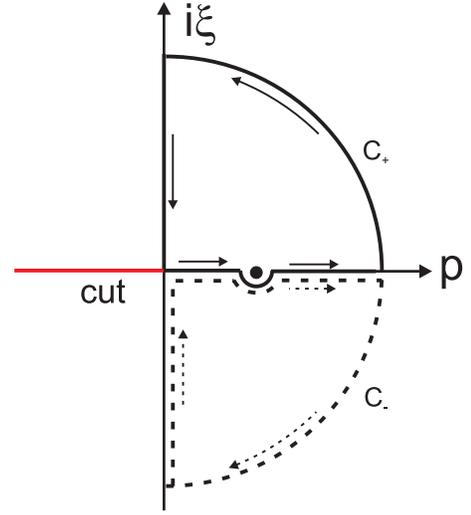}}
\caption{Closed contour in the  complex plane. The pole is shown with a dot.  }
\label{FigContourImp}
\end{figure}

Using  the contours of integration in the complex plane presented in Fig.(\ref{FigContourImp}) 
 for calculations of the first and second  integrals in the right-hand sides in Eq.(\ref{I2}), respectively, one finds
\begin{eqnarray}
I_{1}=I_2^{(\pm)} =-\frac{1}{v}
\sqrt{\frac{p_\varepsilon}{2\pi \hbar R}}e^{i( p_{\varepsilon}R/\hbar+\pi/4)}\nonumber \\
+ i^{3/2}\sqrt{\frac{2 \pi}{R}}\Big[ e^{-i \pi/4 }
\int_0^ {\infty} \frac{\sqrt{\xi}e^{-R\xi}}{i\xi -p_{\varepsilon}}\frac{d\xi}{(2 \pi \hbar)^2} \nonumber \\
+ e^{i \pi/4 }
\int_0^ {\infty} \frac{\sqrt{\xi}e^{-R\xi}}{i\xi +p_{\varepsilon}}\frac{d\xi}{(2 \pi \hbar)^2}\Big]
 \label{I1integr}
\end{eqnarray}

As $Rp_\varepsilon \gg 1$ one may neglect $i \xi$ in the denominators of the integrals  and readily finds
Eq.(\ref{I2final})  of the main text.


\textbf{Calculations of the contour integrals for the "virtual" part of Green's function.\label{AppGVirtual}}

In order to calculate   integrals in Eq.(\ref{VirtIpolar}) it is convenient to use contours  in the upper and the lower complex  planes
 for the first and second integrals respectively as it is shown in
Eq.(\ref{VirtIpolar}) with solid and dotted lines. As a result,    Green's function is presented as follows:
\begin{eqnarray}
G_{s\neq \alpha}(\mathbf{r}^\prime,\mathbf{r})= \sum_{s \neq \alpha} I_s;  \hspace{4.7cm} \nonumber \\
I_s = -\frac{1}{\hbar v} \frac{1}{\sqrt{2\pi \hbar R}}
\hspace{5.5cm} \nonumber \\
\times \Big\{ \sqrt{z_{1}} U_s(z_1,0)e^{i(z_{1}- \pi/4)}
+\sqrt{z_{2}} U_s(z_{2},\pi)e^{-i(z_{2}-\pi/4)}   \Big\}\nonumber \\
+\int_0^{i \infty}\frac{ d z \sqrt{z}}{(2 \pi \hbar)^2}e^{i z R} \Big\{
\frac{U_s(z,0)e^{-i \pi/4}}{\varepsilon- \varepsilon_s(z,0)}
- \frac{U_s(-z,\pi)e^{i \pi/4}}{\varepsilon- \varepsilon_s(-z,\pi)}   \Big\} \nonumber \\
+\int_{C_b^{1}}\frac{ d z \sqrt{z}}{(2 \pi \hbar)^2}
\frac{U_s(z,0)e^{-i \pi/4}}{\varepsilon- \varepsilon_s(z,0)}e^{i z R} \hspace{3.4cm} \nonumber \\
+\int_{C_b^{2}}d \frac{ d z \sqrt{z}}{(2 \pi \hbar)^2}
\frac{U_s(z,\pi)e^{i \pi/4}}{\varepsilon- \varepsilon_s(z,\pi)}e^{-i z R} \hspace{3.3cm}
 \label{VirtIinteg}
\end{eqnarray}
where $z_{1} =p_1 +i \xi_1$ and $z_{2} =p_1 -i |\chi_2|$ (where $\xi_1>0$, $\xi_2<0$) are coordinates of the poles in the
first and second integrals in Eq.(\ref{VirtIpolar}), respectively, the residues of which contribute to the contour integration;
the third term in the right-hand side is the integral along the imaginary axis while the last two terms are integrals
along the pathes around  the cuts (those pathes are marked as $C_b^{1}$ and $C_b^{2}$ in Fig.\ref{FigContourVirtual})

 Using  the inequality $R/a \gg 1 $ one  takes the integrals in Eq.(\ref{VirtIinteg}) and
finds       $G_{s\neq \alpha}(\mathbf{r}^\prime,\mathbf{r})$ written
 by the order of magnitude in Eq.(\ref{VirtGreenFinal}) of the main text.

\section{Matrix elements. \label{AppendMatrixelements}}
Here calculations of  matrix elements  with  Kohn-Lattinger functions are presented    for the sake of convenience.
\begin{eqnarray}
A \equiv\int_{\infty}^{\infty}  \chi_{\alpha^\prime,\mathbf{p}^\prime}^\star(\mathbf{r})\chi_{\alpha,\mathbf{p}}(\mathbf{r})=
\sum_{\mathbf{n}=-\infty}^{+\infty}\nonumber \\
 \times \int_{n_x a_x}^{(n_x+1)a_x}
\int_{n_y a_y}^{(n_y+1)a_y}d\mathbf{r}    e^{i(\mathbf{p-p}^\prime)\mathbf{r}}
u^\star_{\alpha^\prime,0}(\mathbf{r})u_{\alpha,0}(\mathbf{r})
 \label{AppNorman1}
\end{eqnarray}
where  $\mathbf{n}= (n_x, n_y)$  while $n_{x,y} = 0, \pm 1, \pm 2, ...$.

Changing integration variables $\mathbf{r}=\mathbf{r}^\prime + \mathbf{a} $ one finds
\begin{eqnarray}
A =
\sum_{\mathbf{n}=-\infty}^{+\infty}e^{i(\mathbf{p-p}^\prime)\mathbf{a}}
\int_{0}^{(\mathbf{a})}
d\mathbf{r}
u^\star_{\alpha^\prime,0}(\mathbf{r})u_{\alpha,0}(\mathbf{r})
 \label{AppNorman2}
\end{eqnarray}
Here summation is over a unit cell.

 Taking the sum one finally finds the normalization condition for the Kohn-Lattinger functions as follows:
\begin{eqnarray}
\int_{\infty}^{\infty}  \chi_{\alpha^\prime,\mathbf{p}^\prime}^\star(\mathbf{r})\chi_{\alpha,\mathbf{p}}(\mathbf{r})
\frac{d \mathbf{r}}{(2 \pi \hbar)^2}=
\delta_{\alpha,\alpha^\prime}\delta(\mathbf{p-p}^\prime)
 \label{AppNormanFinal}
\end{eqnarray}
where the the normalization condition for the periodic functions
$u_{\alpha,0}(\mathbf{r})=u_{\alpha,0}(\mathbf{r+a})=u_{\alpha,0}(\mathbf{r})$ was used:
\begin{eqnarray}
\int_{0}^{(\mathbf{a})}
u^\star_{\alpha^\prime,0}(\mathbf{r})u_{\alpha,0}(\mathbf{r})\frac{d\mathbf{r}}{a^2}=\delta_{\alpha,\alpha^\prime}
 \label{AppNormPeriodFunction}
\end{eqnarray}

Performing analogous calculations one finds  matrix elements of the velocity operator:
\begin{eqnarray}
\int_{\infty}^{\infty}  \chi_{\alpha,\mathbf{p}}^\star(\mathbf{r})\hat{\mathbf{v} }\chi_{\alpha^\prime,\mathbf{p^\prime}}(\mathbf{r})
\frac{d \mathbf{r}}{(2 \pi \hbar)^2}=\delta(\mathbf{p-p}^\prime) \mathbf{v}_{\alpha,\alpha^\prime}
\label{AppVelocity}
\end{eqnarray}
{}

\section{Calculation of  the integral along the cut for the edge scattering \label{contour}}
Using Eq.(\ref{ComplexGraphIntegralOneVally}) of the main text one writes the integral along the branch cut
 in Fig.\ref{OneVallyFig}
as follows:
\begin{eqnarray}
I^{(cont)}=2 \int_{i p_x} ^{i \infty}\frac{e^{i y \xi/\hbar}}{[\varepsilon-
v\sqrt{p_x^2+\xi^2}}d\xi \nonumber \\
=2i \int_{ p_x} ^{ \infty}\frac{e^{- y \zeta/\hbar}}{[\varepsilon-v\sqrt{p_x^2-\zeta^2}]}d\zeta;
 \label{CotourIntegralOneVally}
\end{eqnarray}

 Changing the variables
$\zeta-q \rightarrow \zeta$ one gets
\begin{eqnarray}
I^{(cont)}=2 i e^{- y p_x/\hbar}\int_{0}^{\infty} \frac{ e^{-y \zeta/ \hbar}}{\varepsilon-
i v\sqrt{\zeta (\zeta +2 p_x)}}d\zeta
 \label{AppgraphIntegral}
\end{eqnarray}

As one sees from Eq.(\ref{AppgraphIntegral}) the main contribution of the integrand  to the
integral is at $\zeta  \lesssim \hbar/y$.  This inequality means that the square root in the
integral denominator is much less than $\varepsilon /v$ (note that $|p_x^{(in)}\  \lesssim
\varepsilon/v$). Therefore, neglecting the term with the square root one easily takes the
integral and finds
\begin{eqnarray}
I^{(cont)}=\frac{2i \hbar  e^{-p_x y/\hbar}}{y  \varepsilon}
 \label{graphB}
\end{eqnarray}

\end{appendix}

\end{document}